# 带有输出端扰动的反馈系统中信息速率的分解

**摘要**：文本研究了带有输出端干扰的线性反馈系统中反馈信道的资源分配问题。通过对反馈信道中信息速率的分解，分析了反馈通信信道的资源分配。可以看出，反馈信道的一部分资源不可避免的用于传输输出端的干扰信号，并且该资源的分配独立于控制器的设计。

# Information Rate Decomposition for Feedback Systems with Output Disturbance.

*Abstract- This technical note considers the problem of resource allocation in linear feedback control system with output disturbance. By decomposing the information rate in the feedback communication channel, the channel resource allocation is thoroughly analyzed. The results show that certain amount of resource is used to transmit the output disturbance and this resource allocation is independent from feedback controller design.*

## I. 引言

本文考虑了用控制器通过闭环通信信道控制线性系统的问题。如图 1 所示。时不变线性系统 P 的输出信号在收到干扰之后，通过线性时不变反馈信道 H 反馈到控制器 K，之后控制器 K 产生控制信号并输入到系统 P。在反馈控制领域，这个是一个传统的反馈模型。几十年来，基于这个反馈系统模型的大量研究取得了丰硕的成果。但是，由于存在系统输出端的噪声，在反馈通信信道中，有多少信道资源用于传输有用的信息，例如，有助于系统的稳定性，至今还是一个开放问题。信息理论是通信领域的基础理论。上个世纪九十年代，有向信息概念的提出使信息论开始应用于因果反馈系统的研究。有向信息是传统互信息的在因果系统中的扩展，并且可以用传统的熵来量化[1][2][3]。近些年来，人们发现有向信息可以用来度量带有无噪声反馈通信信道的信道容量[7]。从本质来讲，无噪声反馈通信系统在某种程度上等价于控制领域中的反馈控制系统。因此，有向信息作为一个桥梁，链接了信息论和反馈控制理论。并且通过信息论的结果，可以更好的理解和研究反馈控制系统中的性能极限[4-6][8-9]。同时，控制理论中的方法和结论也可以用来解决信息论里的理论问题。

在本文中，通过对反馈通信信道中信息速率的分解，得以了解反馈系统中信息流的成分和其各成分的作用。[12-14][18-19] 得到了一些带有外界加载噪声的反馈系统性能极限初步的结果。

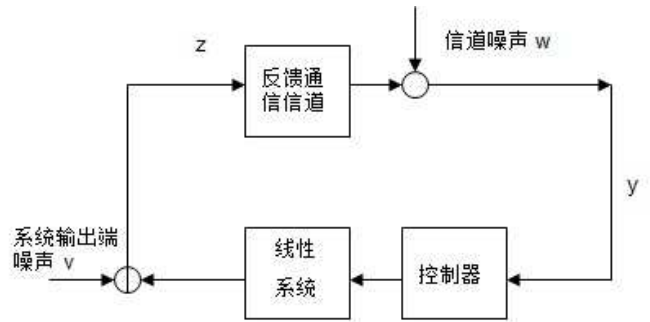

图 1. 带有输出端噪声的线性反馈系统

## II. 有向信息和功率谱密度

首先，本文引入有向信息的概念。有向信息在信息论领域有着广泛的应用。它可以用来量化无噪声反的通信信道的信道容量。对于有噪声反馈的通信信道，有向信息依然可以用来分析信道里的信息流，并且可以用来得到一些有噪声反馈通信信道容量的上界和下界[10-11] [15]。除此之外，有向信息与经济学里的 Granger Causality，数据压缩等都有很直接的联系

[17]，并且可以应用到目标跟踪[16]。有向信息和有向信息速率定义如下。

**定义**：令 X 和 Y 是两个稳定随机过程，它们之间的有向信息和有向信息速率定义为

$$I(X_1^n -> Y_1^n) = \sum_{i=1}^{n} I(Y_i; X_1^n | Y_1^{i-1})$$

$$I_\infty(X -> Y) = \lim_{n->\infty} \frac{I(X_1^n -> Y_1^n)}{n}$$

其中，$X_1^n = \{X_1, X_2, \cdots X_n\}$。通过数学表达式可以看出，有向信息度量了信息流的因果关系。也就是说，序列 Y 是由序列 X 以时间顺序依次产生的。那么从序列 X 到序列 Y 的有向信息表明了从序列 Y 以因果顺序依次含有序列 X 的信息的总和。

接下来，本文引入功率谱密度和反馈控制理论中的敏感度方程。

**定义**：对于一个渐进稳定的随机信号 a，它的功率谱密度表示为 $S_a(e^{jw})$。

功率谱密度定义了信号或者时间序列的功率如何随频率分布。这里功率可能是实际物理上的功率，或者更经常便于表示抽象的信号被定义为信号数值的平方，也就是当信号的负载为 1 欧姆(ohm)时的实际功率。

**定义**：对于给定的两个渐进稳定的随机信号 a 和 b，b 相对于 a 的敏感度方程定义为

$$S_{a,b}(e^{jw}) = \sqrt{\frac{S_a(e^{jw})}{S_b(e^{jw})}}$$

这个敏感度方程并不等于传统反馈控制系统中的敏感度方程。它是传统敏感度方程的一个扩展。

### III. 反馈线性系统的信息速率分解

首先，作为可以度量反馈通信信道中信息速率的有向信息，可以用功率谱密度的形式表达。

**定理 1** 假设一个反馈控制系统如图 1 所示。P 是线性时不变系统，它的初始状态变量表示为 $x_0$。K 是控制器。信号 v 是系统输出端的加载噪声。信号 w 是反馈过程中加载的稳定高斯噪声。反馈信道的有向信息速率可以表达为功率谱密度的形式如下。

$$I_\infty(Z -> Y) = \frac{1}{2\pi} \int_{-\pi}^{\pi} \log(S_{Y,W}(e^{jw})) dw$$

证明：

$$I_\infty(Z -> Y) = h_\infty(Y) - h_\infty(W)$$

$$= \frac{1}{2\pi} \int_{-\pi}^{\pi} \frac{1}{2} \log(2\pi e S_Y(e^{jw})) dw$$

$$- \frac{1}{2\pi} \int_{-\pi}^{\pi} \frac{1}{2} \log(2\pi e S_W(e^{jw})) dw$$

$$= \frac{1}{2\pi} \int_{-\pi}^{\pi} \frac{1}{2} \log(\frac{S_Y(e^{jw})}{S_W(e^{jw})}) dw$$

$$= \frac{1}{2\pi} \int_{-\pi}^{\pi} \log(S_{Y,W}(e^{jw})) dw$$

证明完毕。

接下来的一个重要的问题是，在反馈通信信道中，有多少信道资源用于稳定线性系统，而又有多少资源（速率）用于传递系统输出端的噪声而成为不可避免的资源浪费。下面，通过对表达式

$$\frac{1}{2\pi} \int_{-\pi}^{\pi} \log(S_{Y,W}(e^{jw})) dw$$

的进一步分解，回答了上面的问题。

**定理 2**：假设一个反馈控制系统如图 1 所示。P 是线性时不变系统，它的初始状态变量表示为 $x_0$。K 是控制器。信号 v 是系统输出端的加载噪声。信号 w 是反馈过程中加载的稳定高斯噪声。此外，$F_{wy}$ 表示从噪声 W 到系统输出信号 Y 的闭环传递方程，$F_{vy}$ 表示从系统输出噪声 V 到系统输出信号 Y 的闭环传递方程。则下面的等式是恒成立的，

$$\frac{1}{2\pi}\int_{-\pi}^{\pi}\log(S_{Y,W}(e^{jw}))dw$$

$$=\frac{1}{2\pi}\int_{-\pi}^{\pi}\log(F_{wy}(e^{jw}))dw+$$

$$\frac{1}{2\pi}\int_{-\pi}^{\pi}\frac{1}{2}\log(1+\frac{|F_{vy}(e^{jw})|^2 S_V(e^{jw})}{|F_{wy}(e^{jw})|^2 S_W(e^{jw})})dw$$

$$=\frac{1}{2\pi}\int_{-\pi}^{\pi}\log(S(e^{jw}))dw+$$

$$\frac{1}{2\pi}\int_{-\pi}^{\pi}\frac{1}{2}\log(1+\frac{|H(e^{jw})|^2 S_V(e^{jw})}{S_W(e^{jw})})dw$$

其中，

$$S(e^{jw})=\frac{1}{1-P(e^{jw})K(e^{jw})}$$

是该闭环系统的敏感度方程。如果反馈信道是功率为 $\sigma_w^2$ 的 AWGN，系统输出端噪声是功率为 $\sigma_v^2$ 高斯白噪声，则该等式可以简化为

$$\frac{1}{2\pi}\int_{-\pi}^{\pi}\log(S_{Y,W}(e^{jw}))dw=$$

$$\frac{1}{2\pi}\int_{-\pi}^{\pi}\log(S(e^{jw}))dw+\frac{1}{2}\log(1+\frac{\sigma_v^2}{\sigma_w^2})$$

**证明**：根据定义

$$S_{a,b}(e^{jw})=\sqrt{\frac{S_a(e^{jw})}{S_b(e^{jw})}}$$

可以得出，

$$\frac{1}{2\pi}\int_{-\pi}^{\pi}\log(S_{Y,W}(e^{jw}))dw=$$

$$\frac{1}{2\pi}\int_{-\pi}^{\pi}\frac{1}{2}\log(S_Y(e^{jw}))dw-\frac{1}{2\pi}\int_{-\pi}^{\pi}\frac{1}{2}\log(S_W(e^{jw}))dw$$

在这个闭环反馈系统中，给定控制器，系统输出信号 Y 的频谱是由信道噪声 W 和系统输出端噪声 V 决定的，具体为

$$S_Y(e^{jw})=|F_{wy}(e^{jw})|^2 S_W(e^{jw})+|F_{vy}(e^{jw})|^2 S_V(e^{jw})$$

带入这个等式，可以得到

$$\frac{1}{2\pi}\int_{-\pi}^{\pi}\log(S_{Y,W}(e^{jw}))dw$$

$$=\frac{1}{2\pi}\int_{-\pi}^{\pi}\frac{1}{2}\log(|F_{wy}(e^{jw})|^2 S_W(e^{jw})+|F_{vy}(e^{jw})|^2 S_V(e^{jw}))dw$$

$$-\frac{1}{2\pi}\int_{-\pi}^{\pi}\frac{1}{2}\log(S_W(e^{jw}))dw$$

$$=\frac{1}{2\pi}\int_{-\pi}^{\pi}\frac{1}{2}\log(|F_{wy}(e^{jw})|^2+|F_{vy}(e^{jw})|^2\frac{S_V(e^{jw})}{S_W(e^{jw})})dw$$

$$=\frac{1}{2\pi}\int_{-\pi}^{\pi}\frac{1}{2}\log(|F_{wy}(e^{jw})|^2(1++\frac{|F_{vy}(e^{jw})|^2}{|F_{wy}(e^{jw})|^2}\frac{S_V(e^{jw})}{S_W(e^{jw})}))dw$$

$$=\frac{1}{2\pi}\int_{-\pi}^{\pi}\frac{1}{2}\log|F_{wy}(e^{jw})|^2 dw$$

$$+\frac{1}{2\pi}\int_{-\pi}^{\pi}\frac{1}{2}\log(1+\frac{|F_{vy}(e^{jw})|^2}{|F_{wy}(e^{jw})|^2}\frac{S_V(e^{jw})}{S_W(e^{jw})})dw$$

$$=\frac{1}{2\pi}\int_{-\pi}^{\pi}\log|F_{wy}(e^{jw})|dw$$

$$+\frac{1}{2\pi}\int_{-\pi}^{\pi}\frac{1}{2}\log(1+\frac{|F_{vy}(e^{jw})|^2}{|F_{wy}(e^{jw})|^2}\frac{S_V(e^{jw})}{S_W(e^{jw})})dw$$

接下来，根据线性闭环反馈系统的特性，传递方程可以表示为

$$F_{vy}(e^{jw})=\frac{H(e^{jw})}{1-P(e^{jw})K(e^{jw})}$$

$$F_{wy}(e^{jw})=\frac{1}{1-P(e^{jw})K(e^{jw})}$$

另外，闭环线性系统的敏感度方程为

$$S(e^{jw})=\frac{1}{1-P(e^{jw})K(e^{jw})}$$

基于上面的三个等式，可以得到

$$\frac{1}{2\pi}\int_{-\pi}^{\pi}\log(S_{Y,W}(e^{jw}))dw$$

$$=\frac{1}{2\pi}\int_{-\pi}^{\pi}\log|F_{wy}(e^{jw})|dw$$

$$+\frac{1}{2\pi}\int_{-\pi}^{\pi}\frac{1}{2}\log(1+\frac{|F_{vy}(e^{jw})|^2}{|F_{wy}(e^{jw})|^2}\frac{S_V(e^{jw})}{S_W(e^{jw})})dw$$

$$=\frac{1}{2\pi}\int_{-\pi}^{\pi}\log|S(e^{jw})|dw$$

$$+\frac{1}{2\pi}\int_{-\pi}^{\pi}\frac{1}{2}\log(1+\frac{|\frac{H(e^{jw})}{1-P(e^{jw})K(e^{jw})}|^2}{|\frac{1}{1-P(e^{jw})K(e^{jw})}|^2}\frac{S_V(e^{jw})}{S_W(e^{jw})})dw$$

$$=\frac{1}{2\pi}\int_{-\pi}^{\pi}\log|S(e^{jw})|dw$$

$$+\frac{1}{2\pi}\int_{-\pi}^{\pi}\frac{1}{2}\log(1+|H(e^{jw})|^2\frac{S_V(e^{jw})}{S_W(e^{jw})})dw$$

如果反馈信道是功率为 $\sigma_w^2$ 的 AWGN，系统输出端噪声是功率为 $\sigma_v^2$ 高斯白噪声，则

$$S_W(e^{jw})=\sigma_w^2,\quad S_Y(e^{jw})=\sigma_y^2$$

$$H(e^{jw})=1$$

所以，上面的恒等式可以简化为

$$\frac{1}{2\pi}\int_{-\pi}^{\pi}\log(S_{Y,W}(e^{jw}))dw=$$

$$\frac{1}{2\pi}\int_{-\pi}^{\pi}\log(S(e^{jw}))dw+\frac{1}{2}\log(1+\frac{\sigma_V^2}{\sigma_w^2})$$

证明完毕。

通过上面的定理可以看出，闭环反馈系统的信息理论极限被分解为两个部分。第一个部分

$$\frac{1}{2\pi}\int_{-\pi}^{\pi}\log(S(e^{jw}))dw$$

为标准的 Bode 积分，并且其取值为

$$\frac{1}{2\pi}\int_{-\pi}^{\pi}\log(S(e^{jw}))dw=\sum\max(0,\lambda_i)$$

其中 $\lambda_i$ 为线性系统 P 的极点值。带入上面的取值，可以得到

$$\frac{1}{2\pi}\int_{-\pi}^{\pi}\log(S_{Y,W}(e^{jw}))dw=$$

$$\sum\max(0,\lambda_i)+\frac{1}{2}\log(1+\frac{\sigma_V^2}{\sigma_w^2})$$

这个是 Lemma 4.3 [12] 的结果。第二个部分为

$$\frac{1}{2}\log(1+\frac{\sigma_V^2}{\sigma_w^2})$$

这个部分可以理解为在反馈通信信道中传输系统输出端的噪声，也就是说，系统输出端的噪声被看做为有用的信息在信道中被传输，其传输的功率为 $\sigma_v^2$。在实际的系统中，这个传输所用掉的资源是不可避免的浪费。通过上面的表达式可以看出，这个信道资源的使用独立于控制器的设计。

## IV. 结束语

本文考虑了带有输出端干扰的线性反馈系统。通过对其反馈通信信道中信息速率的分解，深入理解了反馈信道中资源的分配。结论说明对于有输出端噪声的反馈系统，反馈信道中资源的浪费是不可避免的，并且独立于控制器的设计。

## 参考文献

[1] J. Massey, "Causality, feedback and directed information," in Proc. Intl. Symp. Inf. Theory and its Appl., Hawaii, USA, Nov. 1990, pp. 303–305.
[2] J. Massey and P. Massey, "Conservation of mutual and directed information," in Proc. IEEE Int. Symp. Information Theory, Sept. 2005, pp. 157–158
[3] G. Kramer, "Directed information for channels with feedback." Ph.D. dissertation, Swiss federal institute of technology, 1998.
[4] J. Baillieul, "Feedback Designs in Information Based Control," Stochastic Theory and Control: Proceedings of a Workshop Held in Lawrence, Kansas, 2002.
[5] A. Savkin, "Analysis and synthesis of networked control systems: Topological entropy, observability, robustness and optimal control," Automatica, vol. 42, pp. 51–62, 2006.

[6] G. Nair and R. Evans, "Stabilizability of stochastic linear systems with finite feedback data rates," SIAM Journal on Control and Optimization, vol. 43, no. 2, pp. 413–436, 2004.

[7] S. Tatikonda and S. Mitter, "The capacity of channels with feedback," IEEE Trans. Inf. Theory, vol. 55, no. 1, pp. 323–349, Jan. 2009.

[8] S. Tatikonda, A. Sahai, and S. Mitter, "Stochastic linear control over a communication channel," IEEE Transactions on Automatic Control, vol. 49, no. 9, pp. 1549–1561, 2004.

[9] S. Tatikonda and S. Mitter, "Control under communication constraints," IEEE Transactions on Automatic Control, vol. 49, no. 7, pp. 1056–1068, July 2004.

[10] C. Li, and N. Elia. "Bounds on the achievable rate of noisy feedback gaussian channels under linear feedback coding scheme." In Proc. IEEE Int. Symp. Information Theory, 2011, pp 169-173

[11] C. Li and N. Elia. "The information theoretic charaterization of the capacity of channels with noisy feedback." In Proc. IEEE Int. Symp. Information Theory, 2011. pp. 174-178

[12] N. Martins and M. Dahleh, "Feedback control in the presence of noisy channels: "Bode-Like" fundamental limitations of performance," IEEE Trans. Autom. Control, vol. 53, no. 7, pp. 1604–1615, Aug. 2008.

[13] N. Martins, M. Dahleh, and J. Doyle, "Fundamental limitations of disturbance attenuation in the presence of side information," IEEE Transactions on Automatic Control, vol. 52, no. 1, pp. 56–66, 2007.

[14] N. Martins, M. Dahleh, and N. Elia, "Feedback stabilization of uncertain systems in the presence of a direct link," IEEE Transactions on Automatic Control, vol. 51, no. 3, pp. 438–447, 2006.

[15] C. Li, and N. Elia. "Upper bound on the capacity of gaussian channels with noisy feedback." In Communication, Control, and Computing (Allerton), 2011 49[th] Annual Allerton Conference on, pp 84-89.

[16] H. Zhang and Y.-X. Sun, "Directed information and mutual information in linear feedback tracking systems," in Proc. 6-th World Congress on Intelligent Control and Automation, June 2006, pp. 723–727

[17] H. H. Permuter, Y.-H. Kim, and T. Weissman, "Interpretations of directed information in portfolio theory, data compression, and hypothesis testing," IEEE Trans. Inf. Theory, vol. 57, no. 6, pp. 3248–3259, June 2011.

[18] H. Shingin and Y. Ohta, "Disturbance rejection with information constraints: Performance limitations of a scalar system for bounded and Gaussian disturbances," Automatica, vol. 48, pp. 1111–1116, 2012.

[19] J. Freudenberg, R. Middleton, and J. Braslavsky, "Stabilization with disturbance attenuation over a Gaussian channel," in Proceedings of the 46th IEEE Conference on Decision and Control, New Orleans, USA, 2007